\documentclass[final,3p,times,twocolumn]{elsarticle}
 \biboptions{comma,sort&compress}
\usepackage{here}
 \usepackage{graphicx}
  \usepackage{epsfig}

\def\beq{\begin{equation}}
\def\eeq{\end{equation}}
\def\bea{\begin{eqnarray}}
\def\eea{\end{eqnarray}}

\journal{Nuc. Phys. (Proc. Suppl.)}

\begin{document}

\begin{frontmatter}



\title{Jet Production in proton-proton collisions at \boldmath $\sqrt s~=~7$~TeV with the ATLAS experiment}

\author{{\it Chiara Roda} }
\ead{chiara.roda@cern.ch}
\author{ on behalf of the ATLAS Collaboration}
  \address{Universit\`a degli Studi and Istituto di Fisica Nucleare Pisa, \\
  Largo Pontecorvo 3, 56127 - Pisa, Italy}
  


\begin{abstract}
We report on the first measurements done with the ATLAS experiment of the characteristics of energetic jets produced in proton-proton collisions at the center of mass energy of 7 TeV. Jets are reconstructed using the anti-k$_t$ clustering algorithm with distance parameter R=0.6. The kinematic region investigated in this paper corresponds to jets with transverse momentum $p_T>30$ GeV and rapidity  $|y_{jet} |<2.8$. A critical understanding of the jet production is obtained by comparing the data to predictions based on leading-order QCD matrix elements plus parton shower Monte Carlo simulated events. The results shown are obtained on a data sample corresponding to about $1~nb^{-1}$ of integrated luminosity delivered by the Large Hadron Collider.  
\noindent

\end{abstract}


\end{frontmatter}


\section{Introduction}
The last months of 2009 and the beginning of 2010 have taken the LHC experiments into a new era and the rich sample of proton-proton collisions collected in these periods, at a center of mass energy of 900 GeV and 7 TeV, have allowed to fully test the detector performance.  These data prepare the ground to start the path to the more challenging discovery searches. 
In this paper we report the first measurements, done with the ATLAS experiment~\cite{ATLAS}, of the characteristics of energetic jets produced in proton-proton collisions at the center of mass energy ($\sqrt s$) of 7 TeV.  The huge increase of the jet production cross section with respect to Tevatron implies that already with a luminosity of about $1 pb^{-1}$ we will be able to observe jets at the kinematic limit reached by Tevatron~\cite{TevatronResults} ($p_T \simeq 700 GeV$) allowing a first glance at physics at the TeV scale. Moreover jets will be one of the largest sources of background for the searches for new signals. It is therefore important to have a detailed understanding jet production.
\\
ATLAS is a general purpose detector built around the collision region with a structure made of concentric layers consisting of tracking detectors, calorimeters and muon chambers.  The tracking device,  immersed in a 2 Tesla magnetic field, allows to measure tracks in the rapidity range $|\eta| < 2.5$\footnote{The ATLAS reference system is a cartesian right-handed coordinate system, with the nominal collision point at the origin. The azimuthal angle (radians) is measured around the beam axis, and the 
polar angle $\theta$ is measured with respect to the z-axis. The pseudorapidity is defined as $|\eta|= -ln(tan \theta/2)$ and $p_T$ is the track momentum transverse to the beam direction. }. The calorimeter system covers the region $|\eta|<4.9$  and is composed of detectors using different detection techniques in order to exploit the best performance while maintaining a sufficient radiation resistance in each geometrical region. The calorimeter segmentation is such that several jet shower samplings are provided both in the longitudinal and in the transverse direction.
A full description of the ATLAS detector can be found in~\cite{ATLAS}. The events studied in this analysis were collected during the trigger commissioning. Therefore the trigger signal was simply provided by a series of scintillator counters (Minimum Bias Trigger Scintillators - MBTS) located on either side of the interaction point and covering the region $2.09<|\eta|<3.84$. 
\section{Event Selection}
Events are triggered by the requirement that the MBTS recorded one or more counters above threshold on either side.  Only events with stable beam conditions and passing the online data quality criteria are considered. Beam-related backgrounds and cosmic rays are suppressed by requiring that events have at least one reconstructed vertex with a z-position within 10 cm around the origin of the coordinate system. Additional quality criteria are also applied to ensure that jets are not produced by single noisy calorimeter cells or problematic detector regions~\cite{JetQuality}.
\section{Jet Reconstruction and Calibration}
\begin{figure}[hbt]
\centerline{\includegraphics[width=6.cm]{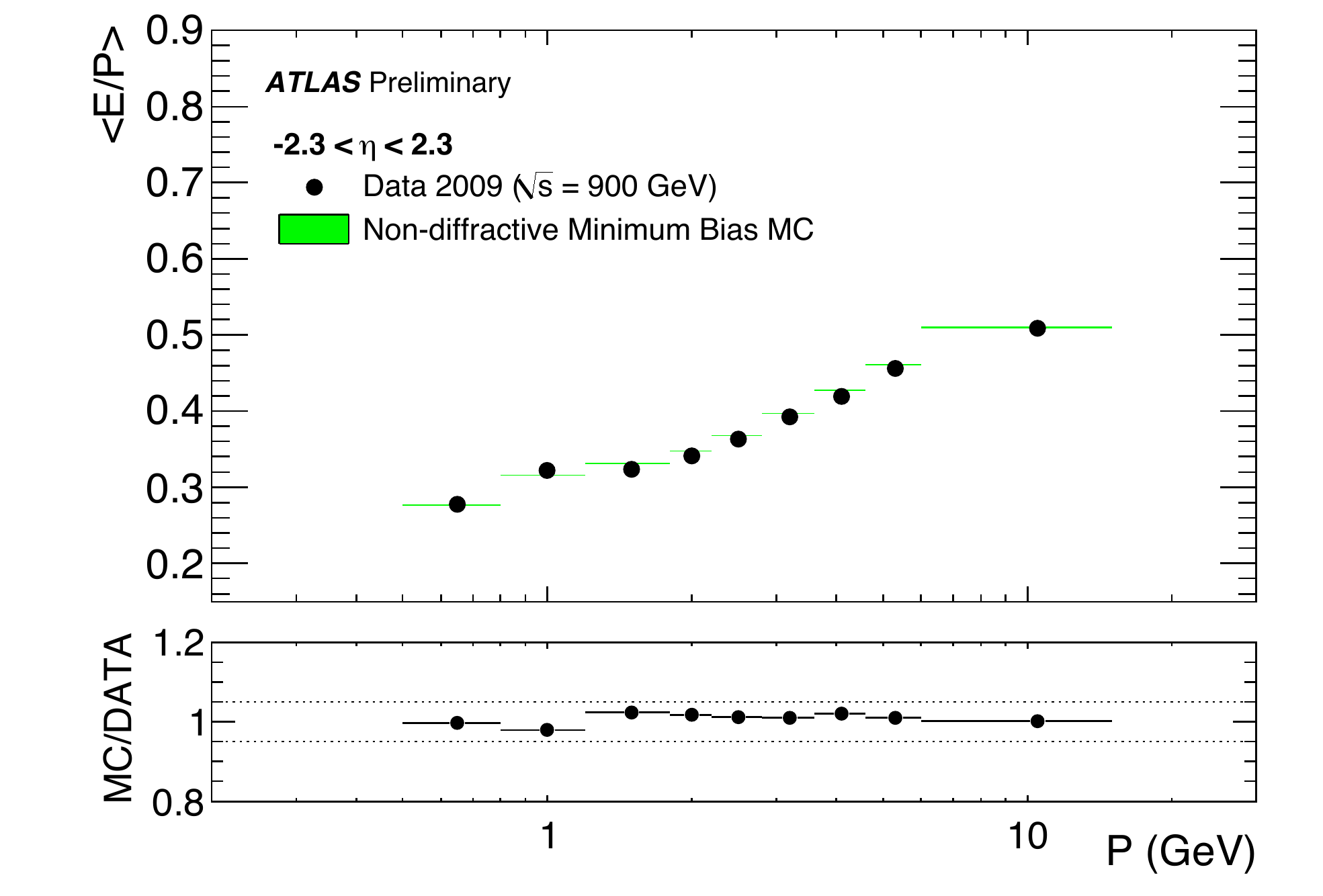}}
\caption{\scriptsize Mean value of E/p as a function of the track momentum for $\eta$=[-2.3, 2.3]. The full circles represent the collision data while the rectangles represent the MC prediction. The lower graph shows the ratio between MC and collision data as a function of the track momentum. }
\label{fig:EOverP} 
\end{figure}

Jets are reconstructed using the anti-k$_t$\cite{AntiKT} algorithm using a distance parameter R=0.6 and full four-momenta recombination. The kinematic region investigated corresponds to jets with transverse momentum $p_T>30$ GeV and rapidity $|y_{jet} |< 2.8$. In this kinematic region the MBTS trigger selects events containing jets with a high efficiency and with a negligible bias. 
\\
Reconstructed jets use as input objects, three dimensional clusters (topological clusters~\cite{CSC-Topological}) built associating calorimeter cells on the basis of the signal-to-noise ratio. Clusters are constructed around cells with a high signal-to-noise ratio and the discrimination to start and expand a cluster is based on the absolute value of the of the signal-to-noise ratio. This algorithm obtains a large noise suppression while introducing a small bias on the cluster energy. Ideally, topological clusters allow to associate together calorimeter signals produced by the same hadron shower. The baseline calibration of the topological clusters is the electromagnetic scale\footnote{The electromagnetic scale is established using test-beam measurements for electrons and muons in the calorimeters.}. At this scale the effect of calorimeter non-compensation or energy losses in un-instrumented material are not corrected for. Therefore, after the jets are identified, a calibration scheme must be applied to correct for these effects and, in general, for any effects that make the reconstructed jet energy different from the reference true jet energy. The reference "truth" jets are obtained by running the anti-k$_t$ algorithm on the ideal final-state of a proton-proton collision where all particles with a lifetime longer than 10~ps are considered stable. This definition includes muons and neutrinos from hadronic decays~\cite{StandardModelHandlesAndCandles}. In ATLAS several calibration schemes~\cite{CSC-Topological} have been developed. The one presently used is the simplest and, while it does not provide the best performance, it allows the most direct evaluation of the systematics. In this scheme the calibration is an average correction factor, obtained from the Monte Carlo (MC) simulation in bins of $\eta$ and $p_T$. The size of the correction ranges from 2 to 1.3 depending on the jet $p_T$ and $\eta$. 
\\
The jet energy scale (JES) systematic is by far the largest uncertainty affecting the jet cross section measurements. One of the largest contribution to the JES uncertainty is given by the energy scale uncertainty of the topological clusters. This can be estimated using the in-situ measurement of the ratio between the energy deposited in the calorimeter (E) over the momentum (p) measured by the tracker for well isolated hadrons. Figure~\ref{fig:EOverP} shows the mean value of the $E/p$ ratio for data and MC simulation in collisions at $\sqrt s = $ 900 GeV~\cite{EOverP-900GeV}.  The ratio between MC and data, as a function of the track momentum, is shown in Figure~\ref{fig:EOverP} (lower curve). Overall, an agreement within 5\% on the average E/p ratio is observed over the range  $\eta = [-2.3,2.3]$ for track momentum ranging from 500 MeV to 10 GeV. 
\\
The same study was carried out on a data sample acquired at $\sqrt s = $7 TeV~\cite{EOverP-7TeV} allowing to extend the track momentum range up to 20 GeV. The results obtained are consistent with those obtained for the 900 GeV sample. For the 7 TeV sample, however, a data-driven procedure must be applied to correct for the neutral background contamination which is not correctly described by the MC. 
\\
The E/p study, integrated with results obtained from the Combined Test Beam~\cite{CTB} and the MC  allows to estimate the energy scale uncertainty of topological clusters. This contributes  3-4\%~\cite{EOverP-7TeV} to the JES uncertainty for jets produced at $0<|\eta|<0.8$ and for $p_T$ ranging from 20 GeV to 1 TeV. 
A more exhaustive estimate of the JES uncertainty, completely based on the MC, has been carried out to determine the contributions given by a long list of sources and to extend the evaluation for the pseudorapidity range $|\eta|<2.8$~\cite{JES}. The resulting JES uncertainty varies between 10\% and 7\% depending on the jet $\eta$ and $p_T$. 
\section{Inclusive Jet Production}
\begin{figure}[hbt] 
   \includegraphics[width=0.22\textwidth]{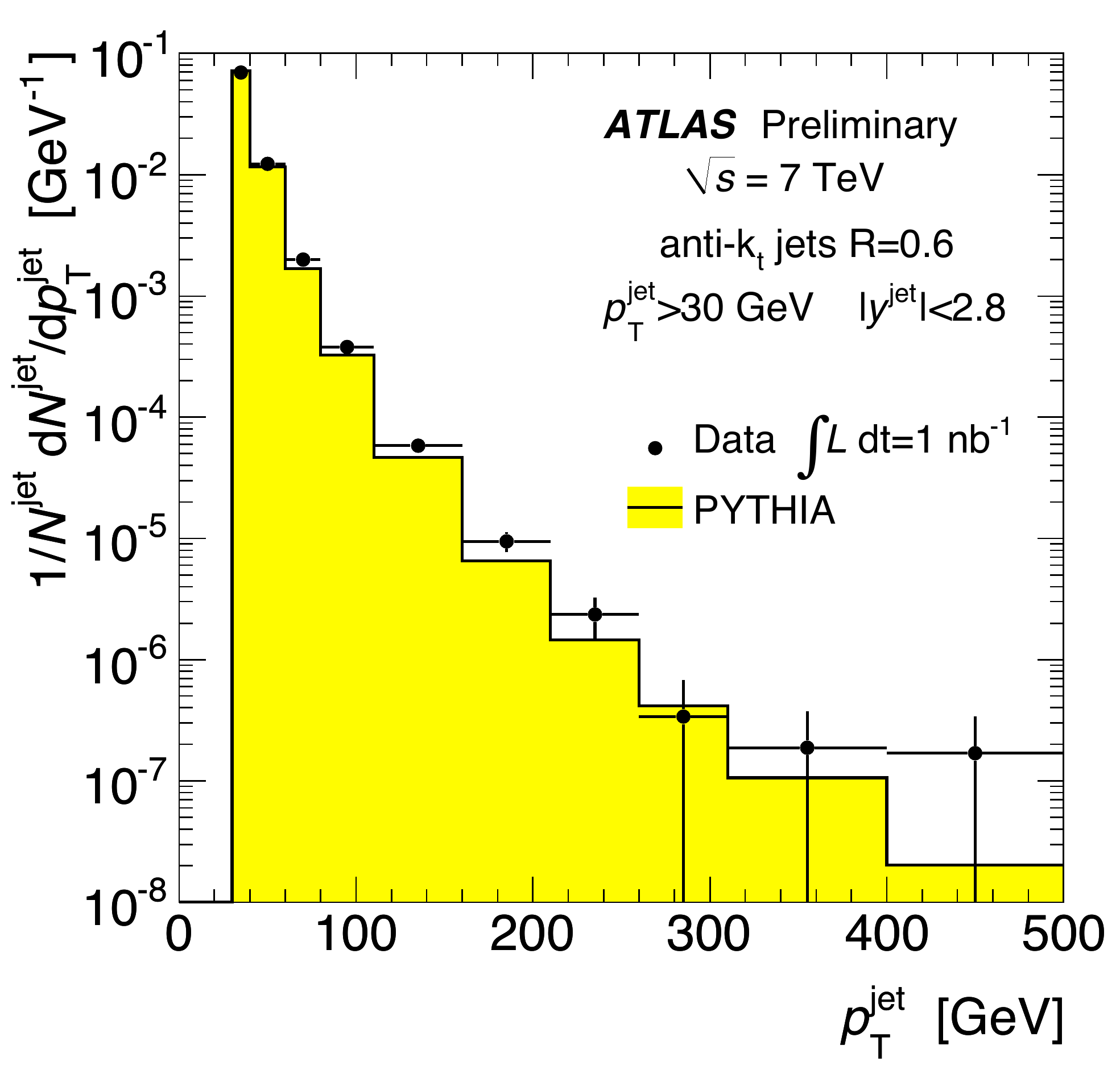}\hfill 
   \includegraphics[width=0.22\textwidth]{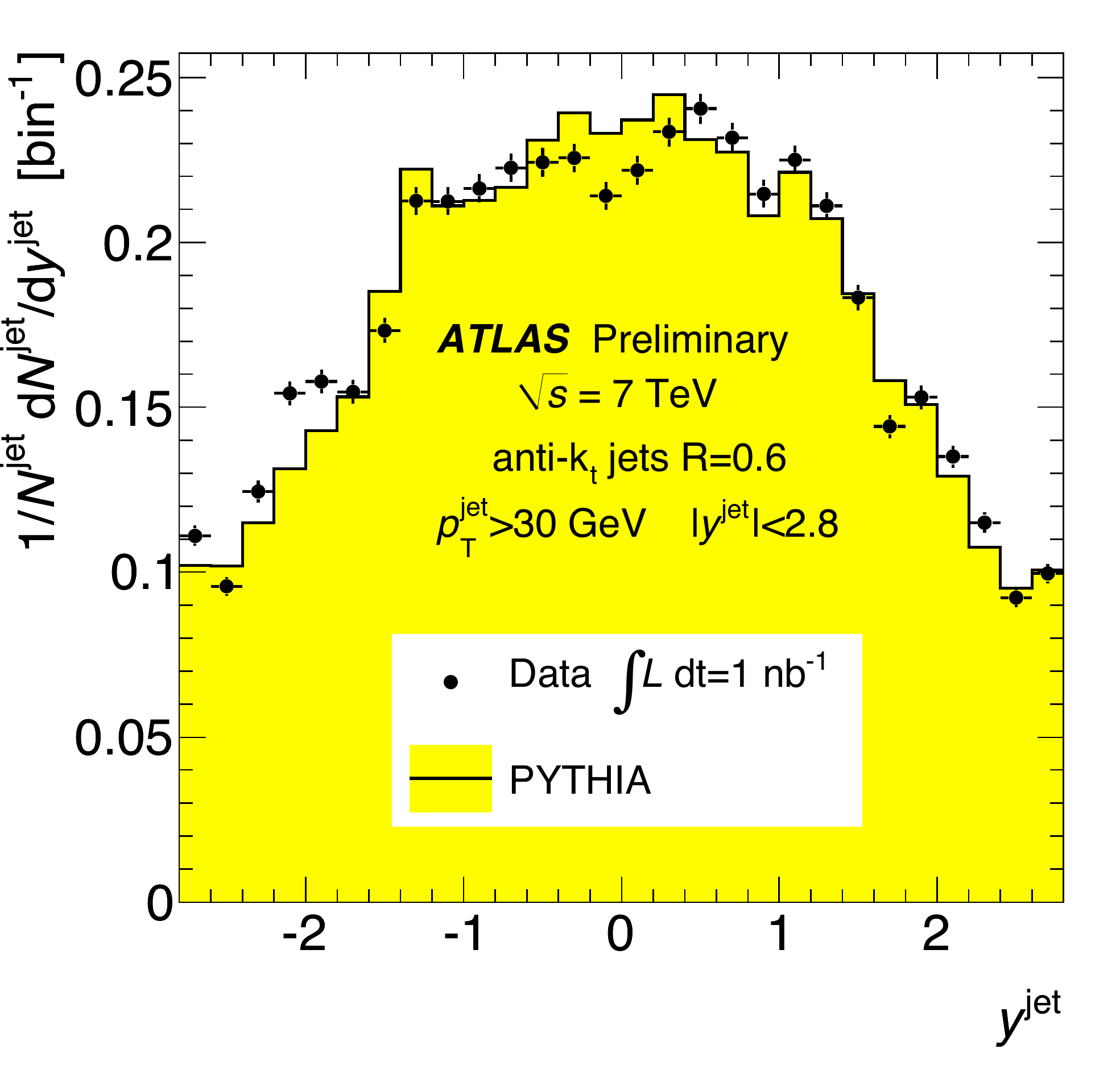} \\
\caption{\scriptsize Observed inclusive $p_T$ (left) and $y_{jet}$ (right) distributions (full circles) for jets with $p_T> 30$ GeV and $|y_{jet} | < 2.8$. Only statistical uncertainties are shown and the distributions are normalized to unit area. The data are compared to PYTHIA Monte Carlo predictions (yellow histograms). }
\label{fig:JetSpectra}
\end{figure}

\begin{figure}[hbt] 
   \includegraphics[width=0.22\textwidth]{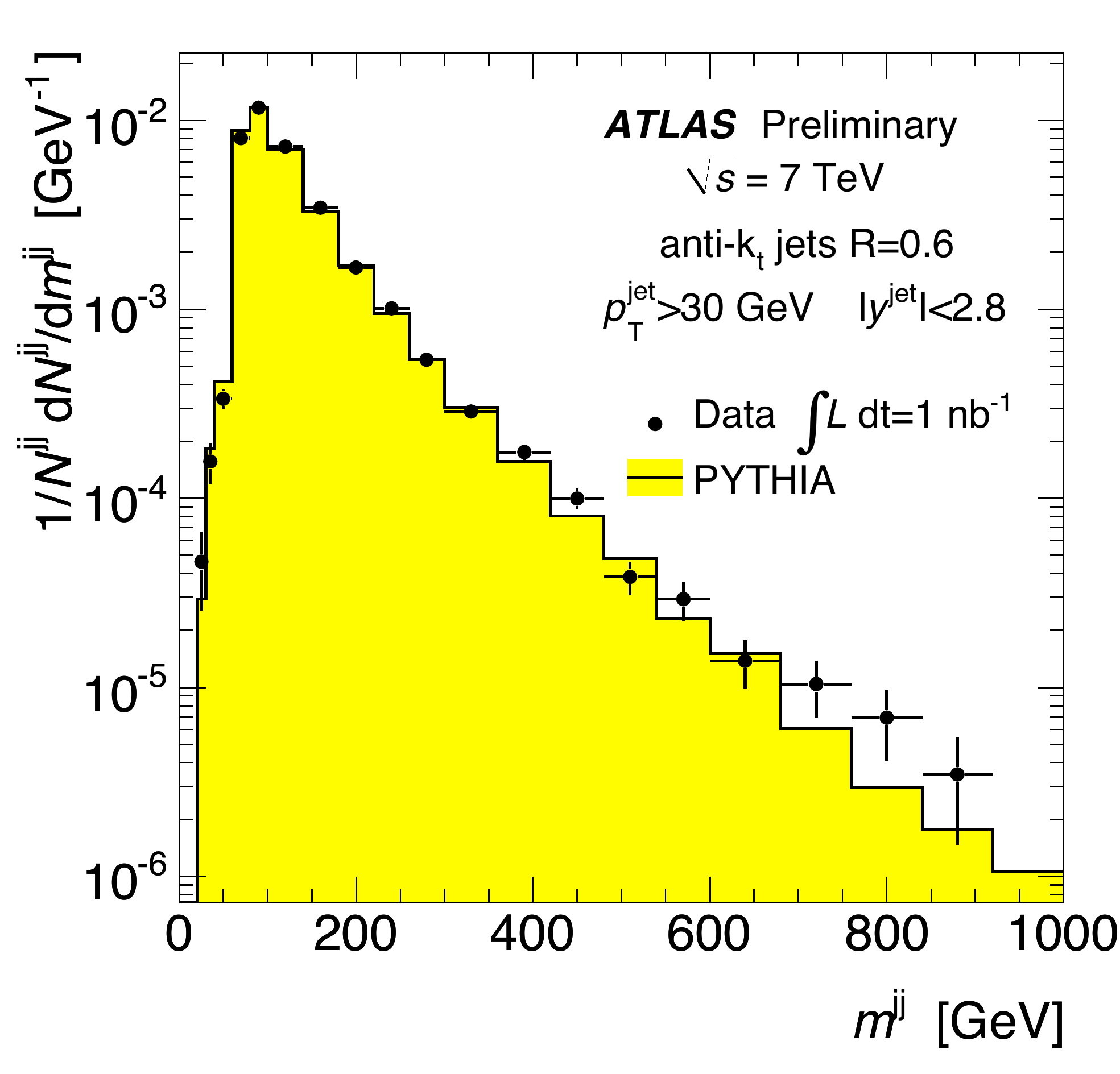}\hfill 
   \includegraphics[width=0.22\textwidth]{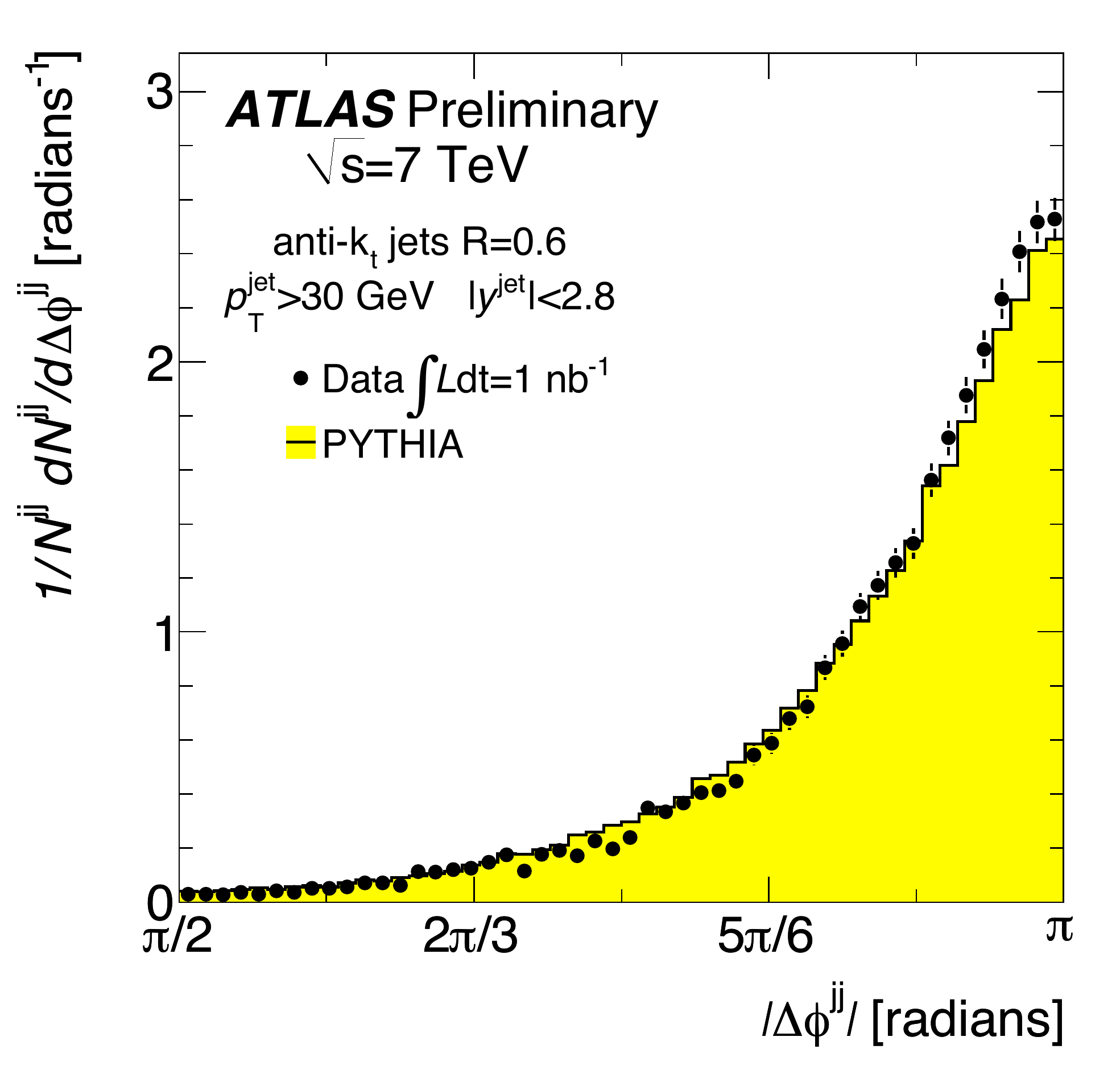} \\
\caption{\scriptsize Observed $m^{jj}$ (left) and $|\Delta \phi^{jj}|$ (right) distributions (black dots) in inclusive dijet events for jets with $p_T>30 GeV$ and $|y_{jet}| < 2.8$. Only statistical uncertainties are included and the distributions are normalized to the total number of di-jet events. The data are compared to PYTHIA Monte Carlo predictions (yellow histograms).  }
\label{fig:DiJetSpectra}
\end{figure}

The results shown~\cite{JetObservation-7TeV} are obtained from a data sample corresponding to about $1~nb^{-1}$ of integrated luminosity. In order to have a critical understanding of jet production, the data are compared to predictions obtained from events generated with PYTHIA 6.4.21, which implements leading-order pQCD matrix elements for the 2 $\rightarrow$ 2 processes plus parton shower in the leading logarithmic approximation. The samples are produced using PYTHIA with MRST LO parton density functions and with a set of parameters  tuned to describe the existing minimum-bias and underlying event data (ATLAS MC09~\cite{MC09}). The generated events are then passed through a full GEANT4~\cite{G4}  simulation of the ATLAS detector. Finally, the MC simulated events are reconstructed and analyzed with the same analysis chain as for the data.\\
The comparison between data and MC is done only for the shapes of the distributions (all distributions are normalized to unit area) at reconstructed level, with no attempt being made to unfold detector effects.  Only statistical errors are shown.
\begin{figure}[hbt] 
\centerline{\includegraphics[width=4.cm]{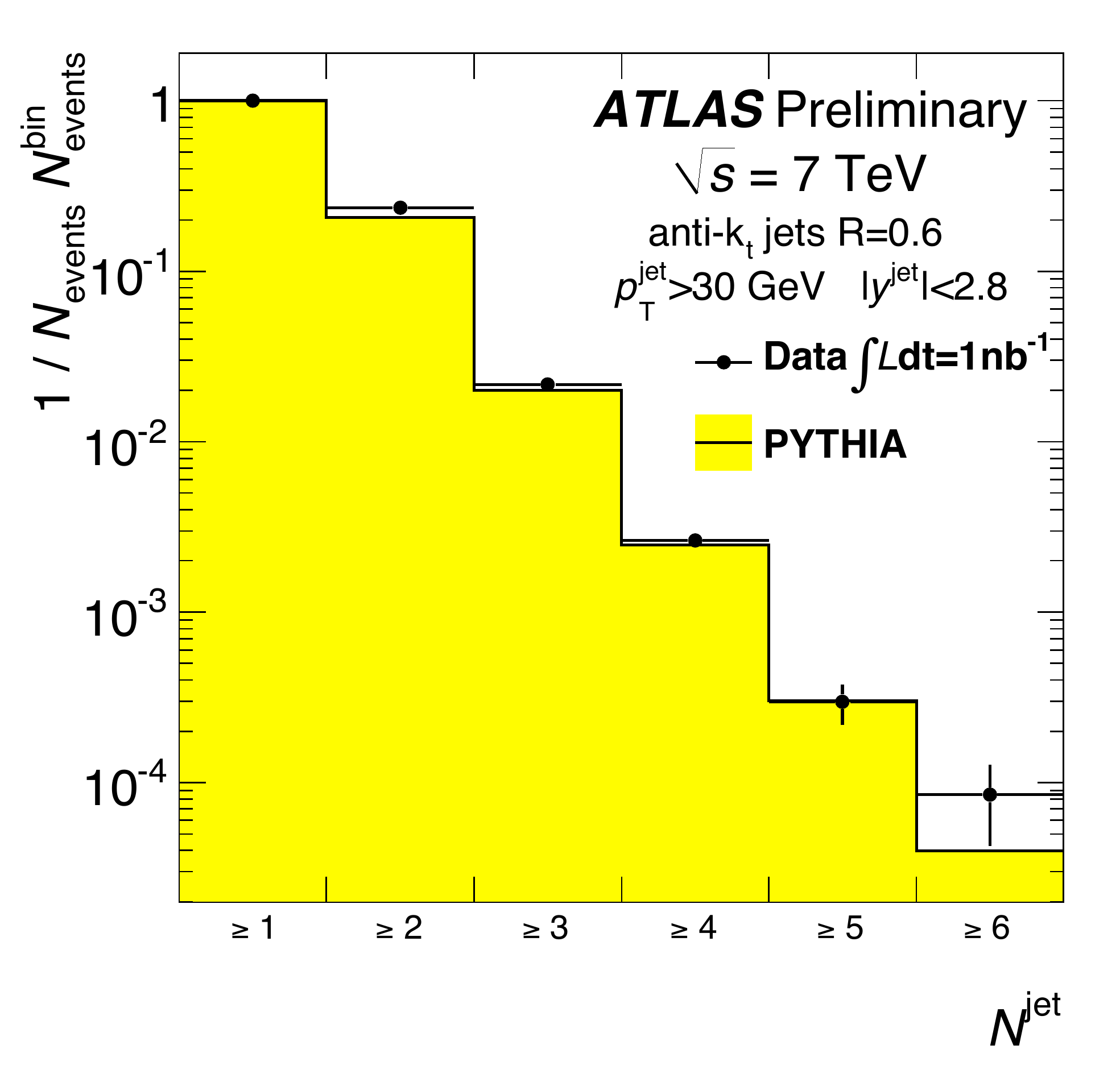}}
\caption{\scriptsize Observed inclusive jet multiplicity ($>=N_{jet}$) distribution (full circles) for jets with $p_T > 30$ GeV and $|y_{jet} | < 2.8$ compared to PYTHIA Monte Carlo prediction (yellow histogram). The distributions are normalized to unity and only statistical uncertainties are shown. }
\label{fig:JetMulti} 
\end{figure} 
\\
The inclusive jet multiplicity for jets with $p_T> 30 GeV$ and $|y_{jet} |< 2.8$ is shown in Figure~\ref{fig:JetMulti} for data (full circles) and MC simulated events (yellow histogram). Events with up to six jets in the final state are observed. The shape of the observed distribution is well described by the MC simulation based on pQCD LO matrix elements plus 
parton shower predictions. The inclusive $p_T$ and $y_{jet}$ distributions are shown in Figure~\ref{fig:JetSpectra} for data (full circles) and MC (yellow histogram). 
The $p_T$ spectrum shows the expected falling shape with increasing $p_T$. Jets with transverse momentum up to 500 GeV are observed. The Monte Carlo simulated events provide a reasonable description of the data even if some small discrepancies are observed in the rapidity distribution.
About 20\% of the selected events contain at least two jets with $p_T > 30$ GeV and $|y_{jet}|<2.8$. The invariant mass $m^{jj}$ of the two leading jets and the azimuthal angular separation $|\Delta \phi^{jj}|$ are shown in Figure~\ref{fig:DiJetSpectra}, compared to PYTHIA MC predictions and normalized using the total number of dijet events. The shape of the $m^{jj}$ distribution at low $m^{jj}$ reflects the limited phase space, as dictated by the thresholds applied on the jet $p_T$ and y. The data show a decreasing $m^{jj}$ spectrum as mjj increases from 50 GeV to about 1 TeV. The shape of the dijet mass spectrum is reasonably described by the MC simulation. 
The observed $|\Delta \phi^{jj}|$ distribution indicates a dominant back-to-back dijet configuration, however the MC prediction tends to underestimate the data at large $|\Delta \phi^{jj}|$.
\section{Jet Shapes}
\begin{figure}[hbt] 
\centerline{\includegraphics[width=0.48\textwidth]{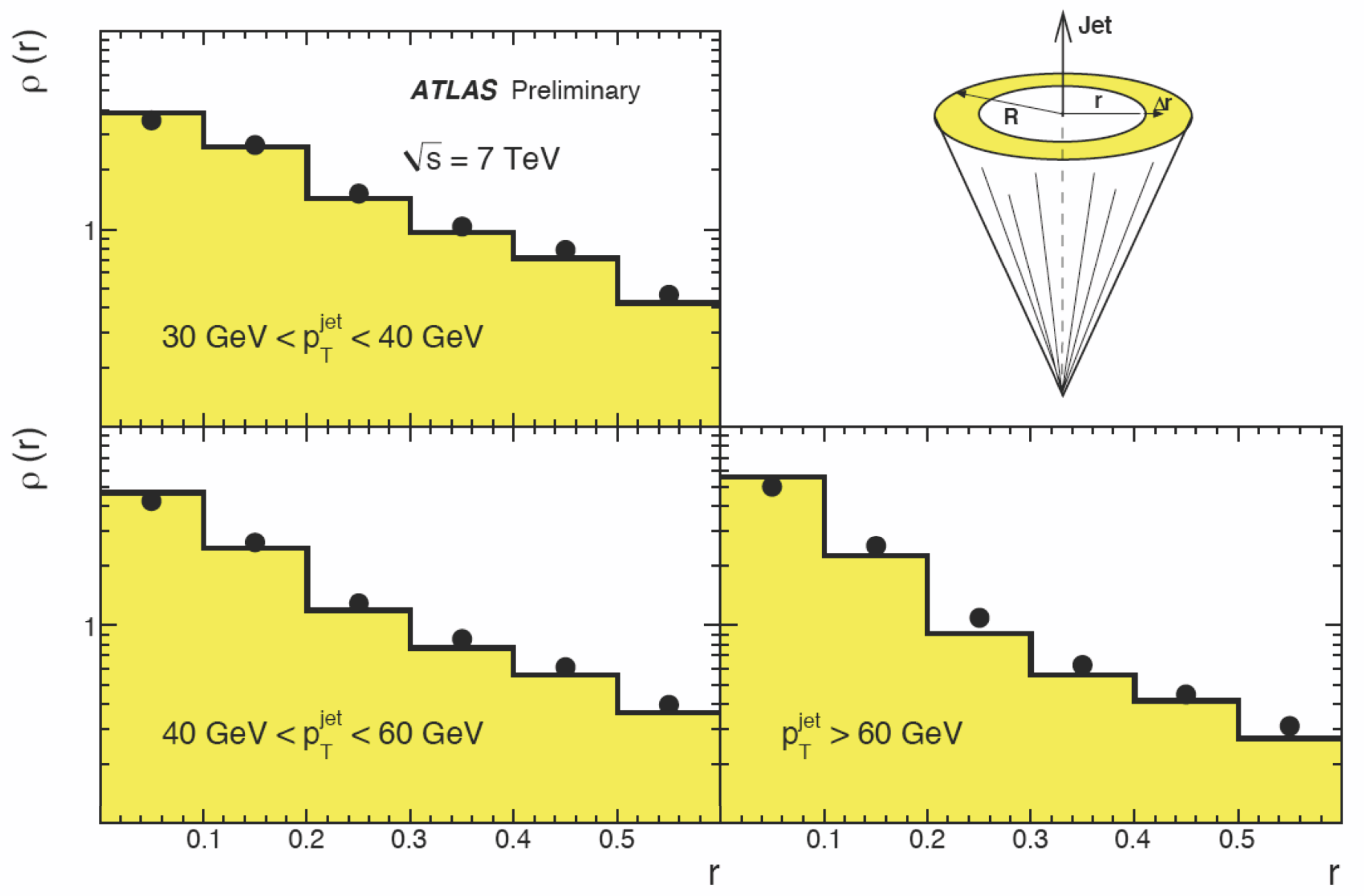}}
\caption{\scriptsize Upper right: Sketch variable used to measure the jet shape as a function of the distance to the jet axis. Upper Left and Lower: Observed differential jet shapes $\rho(r)$ in inclusive jet production for jets with $p_T> 30$ GeV and $|y(jet)| < 2.8$ in different regions of jet transverse momentum for collision data (full circles) and PYTHIA Monte Carlo predictions (yellow histogram).  }
\label{fig:Shape} 
\end{figure} The distribution of the transverse momentum inside the jet cone is studied to verify the presence of collimated flows of particles in the final state. The differential jet shape (see the sketch on the upper right section of Figure~\ref{fig:Shape}) is defined as the average fraction of the jet transverse momentum that lies inside an annulus of inner radius $r-\Delta r/2$ and outer radius $r+\Delta r/2$ around the jet axis, divided by $\Delta r$. The jet transverse momentum is evaluated by the scalar sum of the transverse momentum of the calorimeter clusters in a given annulus. The observed jet shapes in the inclusive jet sample are shown in Figure~\ref{fig:Shape} for different jet momentum ranges. The distributions peak at low r with most of the jet momentum 
concentrated at $r < 0.3$, indicating the presence of a collimated flow of particles around the jet axis. 
The measurements are reasonably well described by the PYTHIA MC simulated events, however the MC tends to produce jets slightly narrower than data.
\section{Conclusions}
We report on the first measurements done by the ATLAS experiment of the characteristics of energetic jets produced in proton-proton collisions  at the unprecedented center of mass energy of 7 TeV. The results shown are based on about 1$~nb^{-1}$ of total luminosity and using the anti-k$_t$ algorithm. The data are compared to Monte Carlo simulated leading-order plus parton shower QCD events. The kinematic region analyzed in this paper includes jets with $p_T>30 GeV$ and $|y_{jet}|<2.8$. The kinematic distributions of the jets in the inclusive-jet and dijet samples show the expected shapes of QCD-driven processes. The study of the jet shapes confirms that the observed jet signal corresponds to collimated flows of particles in the final state. 
\\
More complete results on jet studies will soon become public including jet cross section measurements and we are now ready to work on more challenging calibration schemes that will allow to attain the best performance of the ATLAS calorimeter system.
\section*{Acknowledgements}
C.R would like to thank Stephan Narison and the whole organization committee of the QCD-10 Conference for the warm hospitality and the excellent conference. 












\end{document}